\documentclass[preprint]{aastex}
\newcommand{\sop}{Sol. Phys.}
\newcommand{\jpb}{J. Phys. B: At. Mol. Opt. Phys.}
\newcommand{\adndt}{At. Data and Nucl. Data Tables}
\begin{document}
\title{Electron density diagnostic for hot plasmas in coronal regime by using B-like ions}
\author{G.Y.~Liang\altaffilmark{1,2} and G.~Zhao\altaffilmark{1}}

\email{gyliang@bao.ac.cn}

\altaffiltext{1}{National Astronomical Observatories, CAS, A20
Datun Road, Chaoyang District, Beijing 100012, China.}
\altaffiltext{2}{Department of Physics, University of Strathclyde,
G1 1XQ Glasgow, United Kingdom}

\begin{abstract} Line ratio of $3d-2p$ transition lines
in boron-like spectra of Si~X, S~XII, Ar~XIV and Fe~XXII has been
investigated. Collisional-radiative model calculations reveal that
the line ratio is sensitive to the electron density in ranges of
$n_{\rm e}=4.0\times10^7-3.0\times10^{10}$~cm$^{-3}$,
$4.0\times10^8-3.0\times10^{11}$~cm$^{-3}$,
$3.0\times10^9-4.0\times10^{12}$~cm$^{-3}$ and
$2.0\times10^{12}-3.0\times10^{15}$~cm$^{-3}$, respectively. This
complements the K-shell diagnostics of helium-like ions.  By
comparison between the prediction and the measured values,
effective electron densities in the electron beam ion trap (EBIT)
plasmas performed by Lepson and collaborators at Lawrence
Livermore EBIT, are estimated to be $n_{\rm
e}=3.4^{+0.8}_{-0.6}\times10^{10}$~cm$^{-3}$ and
$5.6^{+1.0}_{-1.1}\times10^{10}$~cm$^{-3}$ for sulphur and argon
plasmas. In case of argon, a good agreement is shown with the
actual electron density derived from N~VI K-shell spectrum. We
further explore the $3d-2p$ transition lines of Si~X and S~XII in
the stellar coronal spectra measured with the Low Energy
Transmission Grating Spectrometer combined with High Resolution
Camera on board the {\it Chandra X-ray Observatory}. The
constrained electron densities show a good agreement with the
those determined from C~V and O~VII K-shell spectra.
\end{abstract}


\section{Introduction}
Electron density ($n_{\rm e}$) plays an important role in fields
of stellar corona, galaxy, interstellar medium of galaxies,
clusters of galaxies, as well as the ground tokamak,
laser-produced, electron beam ion trap hot plasmas. K-shell
spectra (generally refers to the helium-like) shows a powerful
diagnostic potential for the electron density. Since the work of
\cite{GJ69}, which defines a ratio between the {\it forbidden $f$}
line ($1s2s~^3S_1$--$1s^2~^1S_0$) and the {\it intercombination
lines $i$} line ($1s2p~^3P_{2,1}$--$1s^2~^1S_0$), many literatures
reported the ratio by consideration of effects of radiation
field~\citep{NMS01}, dielectronic and radiative
recombinations~\citep{PMD01}. Porquet et al. (2005)
re-investigated the ratio by using updated atomic data of
dielectronic and radiative recombinations.

Since the spectra with high-resolution being available, especially
the launch of {\it Chandra} and {\it XMM-}Newton missions, the
ratio of helium-like spectra has been extensively adopted to
derived the electron density and further estimations of the
stellar structure and the heating mechanism in X-ray emitting
regions~\citep{ABG01,NSB02a,NSB02b,NGS04,SRN05}. Moreover, spatial
information of coronae could be assessed indirectly using
correlation ($EM=n_{\rm e}^2V$, $EM$ refers to an emission
measure) between the electron density and the emission measure.

Besides the K-shell emission lines, many L-shell emission lines
from highly charged oxygen, neon, silicon, sulphur, calcium, iron
and nickel ions have been detected in stellar coronal spectra. For
inactive stars, M-shell transition lines were also
identified~\citep{RMA02}. The diagnostic potential of the L-shell
emission lines has been investigated by many authors, such as the
ratio $I$(50.524)/$I$(50.691) of Si~X, $I$(52.306)/$I$(43.743),
$I$(52.306)/$I$(46.391) of Si~XI addressed in our previous
works~\citep{LZ06a,LZS06b}. Similar characteristics were explored
for Si~VIII, S~X, Ar~XIV and Ca XVI
etc.~\citep{KCF93,KPM00,KAW01,KAK03,LZZ06c}.

For boron-like extreme ultraviolet (EUV) spectra from transitions
of $n=2$--2, \cite{KST00,KKR03} studied density-sensitive line
ratios of Si~X and Ar~XIV by using accurate excitation data from
$R-$matrix method. For $3d-2p$ transition lines spanned in the
soft X-ray region, line intensity ratios of highly charged silicon
ions were investigated in our previous work~\citep{LZS06b}, which
shows a $n_{\rm e}$-sensitivity in the range of $n_{\rm
e}=4.0\times10^7-3.0\times10^{10}$~cm$^{-3}$. By laboratory
measurements in the electron beam ion trap (EBIT), \cite{CBH04}
test the density-sensitive line ratios of Ar~XIV and Fe~XXII in
ranges of $n_{\rm e}=2.0\times10^{10}-1.0\times10^{12}$~cm$^{-3}$
and low-density limit around 1.0$\times10^{12}$~cm$^{-3}$,
respectively. In that work, the actual electron density was
constrained from K-shell N VI spectrum. However, the diagnostic
application of emission lines of S~XII have not been investigated
to our best knowledge, though its two lines at 36.398 and
36.564~\AA\, have been identified for inactive star---Procyon.

In this work, we make a systematically analyses for the $3d-2p$
density-sensitive transition lines of boron-like spectra including
Si~X, S~XII, Ar~XIV and Fe~XXII based on the collisional-radiative
model as given in Sec.~2. In Sec.~3, we explore the $3d-2p$
transition lines of S~XII at 36.398 and 36.564~\AA\, in the
stellar coronal spectra observed by the low energy transmission
grating (LETG) spectrometer on board the {\it Chandra X-ray
observatory}. Furthermore, the electron densities of stellar
corona and laboratory plasmas produced in the EBIT are estimated
as illustrated in Sec.~4, followed by conclusions outlined in
Sec.~5.

\section{Theory}
\subsection{$3d$--$2p$ transition lines} The line intensity ratio
between the $3d~^2D_{5/2}$--$2p~^2P_{3/2}$ transition and the
$3d~^2D_{3/2}$--$2p~^2P_{1/2}$ transition lines are known to be
sensitive to the electron density but on the other hand rather
insensitive to the electron temperature. This feature follows from
the fact that the upper level is populated mainly by collisional
excitation from the density-sensitive lower metastable level. In
the following paragraph, we take S~XII to explain this property.

\begin{figure*}[htb!]
\centering
\includegraphics[angle=0,width=14cm,clip]{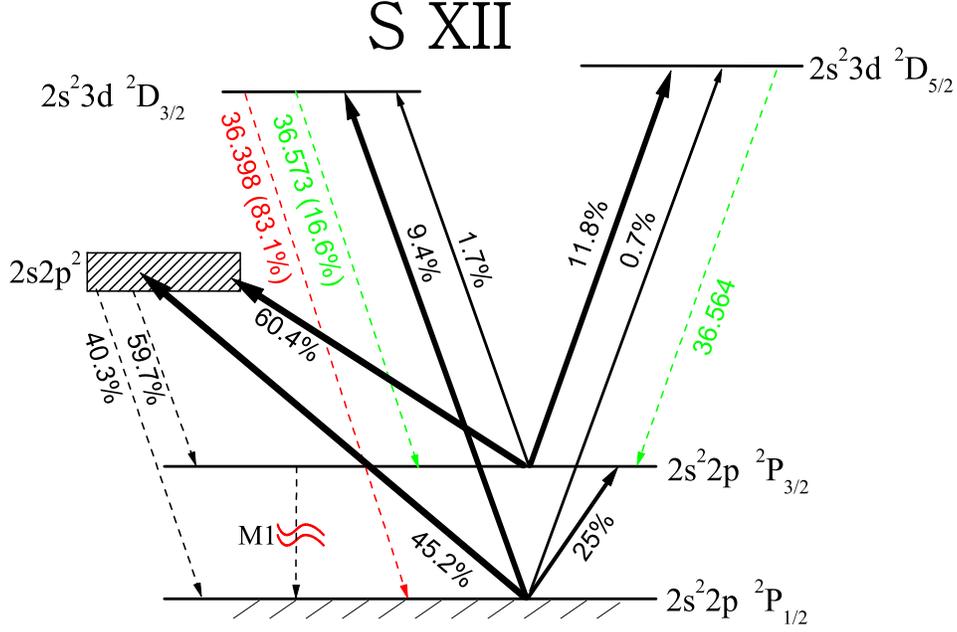}
\caption[short title]{(Color online) Scheme of the processes
responsible for the sensitivity of the $n=3\to n=2$ transitions
(see red- and green-dashed lines) in S~XII against the electron
density. Wavelengths (in \AA\,) and radiative branching fractions
different from 100\% are indicated for the radiative transitions
(dashed lines). Relative magnitudes of collisional excitation
strength are given in percent as well as represented by the
thickness of the corresponding solid lines.} \label{leveldiagram}
\end{figure*}

As shown in Fig.~\ref{leveldiagram}, the levels of the
configuration $2s2p^2$ have the largest excitation rate from the
levels of the ground configuration, whereas the decaying lines
span in the EUV wavelength region. Here we pay special attention
on the X-ray emission lines. The $3d~^2D_{5/2}$ level has a high
excitation rate from and decay rate to the metastable
$2p~^2P_{3/2}$ level, which is forbidden to decay to the ground
state. By increasing the electron density, a significant fraction
of the $^2P_{3/2}$ population can be re-excited to the $^2D_{5/2}$
state, from where it relaxes to the $^2P_{3/2}$ state and,
therefore, the line intensity at 36.564~\AA\, increases. On the
other hand, by the depopulation of the ground state $^2P_{1/2}$
due to the shelving of electrons in the metastable level, the
excitation of the resonance line ($3d~^2D_{3/2}$--$2p~^2P_{1/2}$)
at 36.398~\AA\, of S~XII decreases at high densities.

It is clear that the $2s3d~^2D_{3/2}$ level can also decay to the
metastable $^2P_{3/2}$ level with a low branching ratio. However,
as this tranistion line (at 36.573~\AA\, for S~XII) can not be
resolved from the 36.564~\AA\, line due to the limited spectral
resolution in the present space missions, we have to take into
account this blend effect. Since the $3d~^2D_{3/2}$ is directly
populated from the ground state, the intensity of 36.573~\AA\, has
a similar dependence on $n_{\rm e}$ as the 36.398~\AA\, resonance
line and it relative contribution to the blended 36.564~\AA\, line
becomes dominant at low electron density.

\subsection{Calculation of the line ratio}
Adopting a collisional-radiative model, we calculated the line
ratio [{\it I}($3d_{5/2}-2p_{3/2}$)+{\it
I}($3d_{3/2}-2p_{3/2}$)]/{\it I}($3d_{3/2}-2p_{1/2}$) of Si~X,
S~XII, Ar~XIV and Fe~XXII, as a function of the electron density
for a Maxwellian electron distribution at temperatures of peak
fraction of these ions in the work of~\cite{BBG06}. The transition
wavelengths are listed in Table~\ref{wavelength}. For Si~X and
Ar~XIV, the atomic data including energy levels, radiative decay
rates and impact excitation rates are from our detailed
calculation. Some excitation data within $n=2-2$ transitions are
replaced by more accurate calculation with $R-$matrix method as
explained in previous literatures~\citep{LZS06b,LZZ06c}. For S~XII
and Fe~XXII, we adopt the CHIANTI package of the new version~5.2.
In this code, the atomic data is updated recently as depicted
by~\cite{LZY06}. For S XII, the electron impact excitation data
of~\cite{ZGP94} has been adopted, in which small errors found by
later have been corrected by authors in the present version of the
CHIANTI. Resonance effects for excitations among the lowest 15
fine-structure levels have been included by the {\it R}-matrix
method. For Fe XXII, the electron excitation data between the
lowest 204 levels are from the work of~\cite{BGM01} who adopted
close-coupling {\it R}-matrix method and in conjunction with
intermediate-coupling frame transformation method. The electron
excitation data between higher 205-513 levels adopts the
calculation of~\cite{LG06}, in which the resonance effects were
considered by an isolated-resonance approximation. Above the
ionization energy, the calculation of~\cite{BMK04} with the
Breit-Pauli {\it R}-matrix method was used. Additionally, proton
excitations between the fine-structure states of the ground
configuration and $2s2p^2~^4P$ term were also
included~\citep{FKR97} in the CHIANTI. In this work, close-coupled
impact-parameter method was used.

Figure~\ref{lineratio} shows the prediction of the line ratio at
temperatures of log$T_{\rm e}$=6.1~K for Si~X, 6.3~K for S~XII,
6.5~K for Ar~XIV and 7.1~K for Fe~XXII. This demonstrates that the
ratio is sensitive to the electron density in ranges $n_{\rm
e}=4.0\times10^7-3.0\times10^{10}$~cm$^{-3}$,
$4.0\times10^8-3.0\times10^{11}$~cm$^{-3}$,
$3.0\times10^9-4.0\times10^{12}$~cm$^{-3}$ and
$2.0\times10^{12}-3.0\times10^{15}$~cm$^{-3}$ for Si~X, S~XII,
Ar~XIV and Fe~XXII, respectively. We further calculate the ratio
in the temperature ranges log$T_{\rm e}$=6.0--6.2~K for Si~X,
6.1--6.4~K for S~XII, 6.4--6.6~K for Ar~XIV and 7.0--7.2~K for
Fe~XXII, as illustrated by hatched regions in
Fig.~\ref{lineratio}, which reveals that the ratio is insensitive
to the electron temperature. This indicates that the ratio [{\it
I}($3d_{5/2}-2p_{3/2}$)+{\it I}($3d_{3/2}-2p_{3/2}$)]/{\it
I}($3d_{3/2}-2p_{1/2}$) is a good $n_{\rm e}$-diagnostic method
for hot plasmas, and compensates the K-shell spectra.

\section{Observations}
The wavelength range spanned by the $3d-2p$ transition lines of
the present interested boron-like ions, is covered by the High and
Medium Energy Grating (HEG and MEG) spectrometers, the Low Energy
Transmission Grating (LETG) spectrometer on board {\it Chandra},
as well as the Reflection Grating Spectrometer (RGS) on board {\it
XMM}-Newton with high-resolution. One specific advantage of LETG
observations results from its large wavelength coverage in one
spectrum, that is the $3d-2p$ lines (see Table~\ref{wavelength})
of the interested ions can be detected at the same time.

\begin{figure*}[htb!]
\centering
\includegraphics[angle=0,width=14cm,clip]{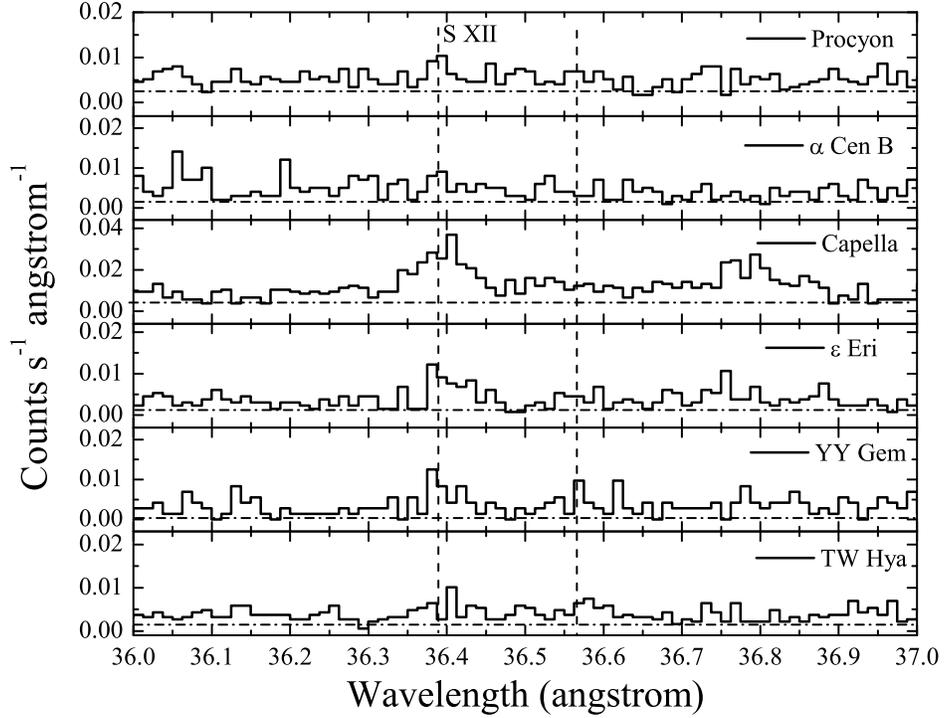}
\caption[short title]{Extracted LETGS spectra of our sample in the
wavelength range 36.0--37.0~\AA\,. The HRC-S instrument was used
for all observation. The prominent lines from S~XII are labelled
by dashed vertical lines. The dash-dotted horizontal lines refer
to the continuum levels for each star.} \label{spectra}
\end{figure*}

Our sample consists of 6 stars including two normal dwarf stars,
i.e., Procyon and $\alpha$~Cen~B, an active late-type dwarf star,
$\epsilon$ Eri, two active binary systems, Capella and YY~Gem, and
an pre-main sequence late-type star--TW~Hya. The properties of our
sample, along with ObsID and exposure time are summarized in
Table~\ref{stars}. All observations adopt grating of LETG combined
with High Resolution Camera (HRC) instrument on board {\it
Chandra} Observatory. In case of Capella, an additional
observation (with ObsID=55) with Advanced CCD Imaging Spectrometer
(ACIS)-S instrument is used. Because ACIS-S instrument has
significant energy resolution to separate overlapping spectral
orders, LETG+ACIS-S observations are better choice for
determination of the electron density. However only one
observation is available from {\it Chandra Data Archival Center}
for our sample. Another goal of analysis for the ACIS-S spectrum
of Capella is to validate whether there is significant
contamination from high-order (refer to $m\geq2$) spectra around
the selected lines (36.398 and 36.564~\AA\,). If line fluxes
derived from the two different observations are comparable, no
contamination from high-order spectra can be concluded. Reduction
of the LETG datasets uses CIAO3.3 software with the science
threads for LETGS/HRC-S observations. Figure~\ref{spectra} shows
the spectra for Procyon, $\alpha$~Cen~B, Capella, $\epsilon$~Eri,
YY~Gem and TW~Hya in the wavelength range of 36.0--37.0~\AA\,.

The $3d-2p$ transition lines of Si~X have been identified and
explained in detail by~\cite{LZ06a}. Argon is an underabundant
element in stellar corona, and no emission lines of Ar~XIV have
been identified so far. Iron is an abundant elements in
astrophysical environment. However, the $3d-2p$ resonance line at
11.769~\AA\, is contaminated by the Fe~XXIII line at 11.738~\AA\,
for Capella~\citep{BCK01}. Here, we pay a special attention on the
identification of $3d-2p$ lines of S~XII at 36.398~\AA\, and
36.564~\AA\, as shown in Fig.~\ref{spectra}.

\section{Results and Discussions}
\subsection{line fluxes of $3d-2p$ transitions of S~XII}
Line fluxes are determined by modelling the spectra locally with
narrow Gaussian profiles and constant value representing
background and (pseudo-)continuum emissions determined in
line-free region. The observed line width is about 0.06~\AA\, over
the interested region, which is comparable with the broadening of
instrument for point-like source. The fluxes have been obtained
after correction for the effective area as listed in
Table~\ref{lineflux}.

For Procyon, the two transition lines have been identified
by~\cite{RMA02} in the RGS and LETG observations. Here the line
fluxes are derived separately again for the co-added LETG
observations, which shows a good agreement with results of
\cite{RMA02} within 1$\sigma$ error as illustrated in
Table~\ref{lineflux}. For other stars, no reports of the
identifications for the two lines can be found to our best
knowledge.

\begin{figure*}[htb!]
\centering
\includegraphics[angle=0,width=14cm,clip]{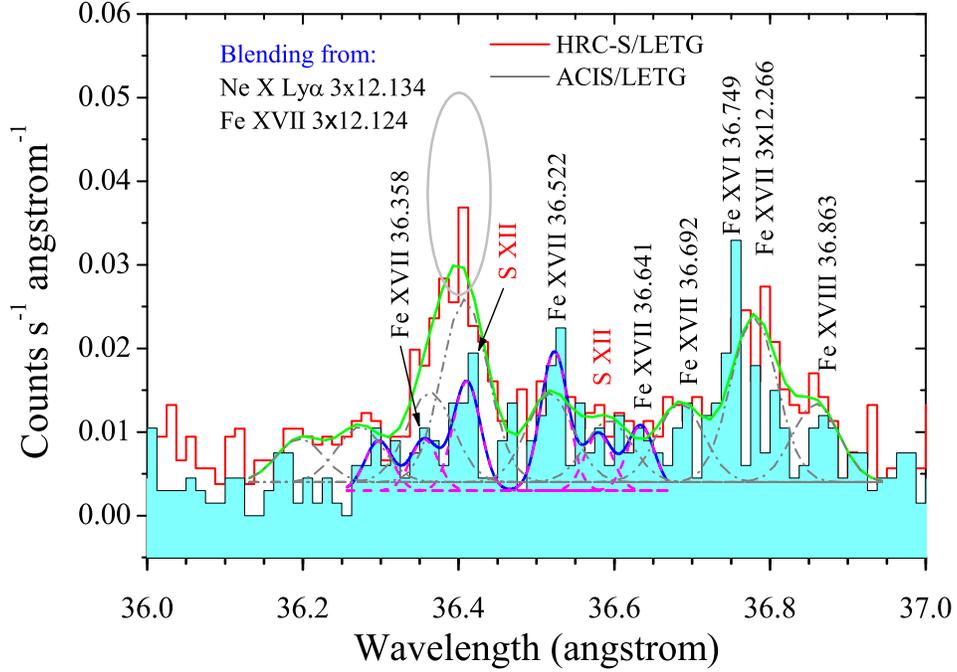}
\caption[short title]{(Color online) Observed spectra (step lines)
of Capella in wavelength range of 36.0---37.0~\AA\, and their
Gaussian fittings (smooth lines) around S~XII lines (marked by red
labels) from $2s^23d~^2D_{3/2}$--$2s^22p~^2P_{1/2}$ transition at
36.398~\AA\, and $2s^23d~^2D_{3/2,5/2}$--$2s^22p~^2P_{3/2}$
transitions at 36.574 and 36.564~\AA\,. Blending from lines of
higher-order diffractions are resolved out by the ACIS instrument
(see black-step curve with filled area), whereas it can not be
resolved out by the HRC-S instrument (see red-step curve). }
\label{ACISspectra}
\end{figure*}

For Capella, we fit the HRC-S and ACIS-S observations separately
with multi-Gaussian components as detailed illustration in
Fig.~\ref{ACISspectra}. It is clear that the line-width in HRC-S
observation is wider than ACIS-S observation, which is due to its
lower resolution. Additionally, for the line at 36.398~\AA\,, the
derived line flux from HRC-S observation is higher than the value
from ACIS-S observation by a factor of $\sim$2.8 (see
Table~\ref{lineflux}). This is due to the contamination from
third-order diffraction of emission line at $\sim$12.136~\AA\,
with flux of 11.04$\pm0.30\times10^{-4}$~photons~cm$^{-2}$s$^{-1}$
(see 7-th column in Table~\ref{lineflux}), resulting from Ne~X
Ly$\alpha$ (12.134~\AA\,) and Fe~XVII (12.124~\AA\,). According to
the line fluxes around 36.394~\AA\, derived from HRC and ACIS
observations as well as the line flux around 12.136~\AA\,, we
conclude that the efficiency of third-order diffraction is about
(17.4$\pm$8.2)\% at the local wavelength range. The different
background levels are due to the different effective areas (6.24
and 8.43~cm$^2$, respectively). A search from the APEC/APED
v1.3.1~\citep{SBL01} indicates that another emission line from
Fe~XVII at 36.358~\AA\, partially blends in the HRC-S observation.
For the emission line around 36.564~\AA\,, the fitting result in
the HRC-S observation is also higher than the value derived from
ACIS-S observation by a factor of $\sim$1.6. This is due to the
blending from Fe~XVII line at 36.692~\AA\, as resolved in case of
ACIS-S observation. Here a question appears that whether there is
a similar blending effect from high-order diffraction for inactive
stars such as Procyon and $\alpha$~Cen~B. Analysis indicates that
no emission line is detected at $\sim$12.136~\AA\, for the
inactive stars, that is, there is no contamination from
third-order diffraction at the emission line at 36.394~\AA\,.

For other three stars including $\epsilon$~Eri, YY~Gem and TW~Hya,
similar procedure is used. The resulting line fluxes are listed in
Table~\ref{lineflux}. However the blending effect has not been
extracted. The line flux of the first order diffraction at
12.136~\AA\, is listed in the 7-th column of Table~\ref{lineflux},
which will be used for the extraction of its contribution at the
third-order diffraction position.

\subsection{$n_{\rm e}$-diagnostic from the ratio}
\begin{figure*}[htb!]
\centering
\includegraphics[angle=0,width=14cm,clip]{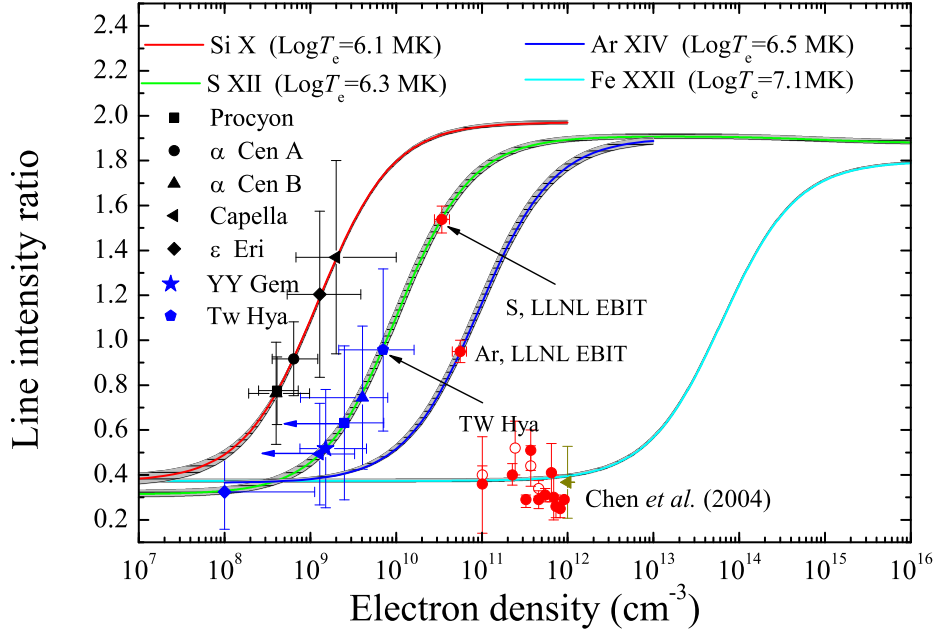}
\caption[short title]{(Color online) Line intensity ratio, [{\it
I}($3d_{5/2}-2p_{3/2}$)+{\it I}($3d_{3/2}-2p_{3/2}$)]/{\it
I}($3d_{3/2}-2p_{1/2}$) of Si~X, S~XII, Ar~XIV and Fe~XXII, as a
function of the electron density for a Maxwellian electron
distribution at temperatures of peak fraction of these ions in the
work of~\cite{BBG06} (see the legend). The hatched areas indicate
predictions for thermal plasmas in the temperature range (in
log$T_{\rm e}$) of 6.0--6.3~K for Si~X, 6.2--6.4~K for S~XII,
6.4--6.6~K for Ar~XIV and 7.0--7.2~K for Fe~XXII. Black symbols
with error bars are the observed line ratios of Si~X in the
stellar corona. Blue symbols with error bars are the observed line
ratios of S~XII in the stellar corona. The red-square symbols with
error bars are the laboratory measured ratios of S~XII and Ar~XIV
in the works of Lepson et al. (2003, 2005). The red-circle symbols
in range of $n_{\rm e}$=$10^{11}-10^{12}$~cm$^{-3}$ are extracted
from the work of Chen et al. (2004), which represent the measured
ratios and electron densities in the EBIT experiments, whereas the
dark-yellow trangile symbol denotes the ratio of Fe~XXII lines
extracted from ACIS-S observation for Capella. } \label{lineratio}
\end{figure*}

The diagnostic results of $n_{\rm e}$ from Si~X for stellar corona
have been reported in our previous work~\citep{LZS06b}. Here the
observed ratio and the constrained electron density with 1$\sigma$
errors are overlayed again in Fig.~\ref{lineratio} (see black
symbols). The comparison with the diagnosed electron density from
K-shell C~V ions, shows a good consistency, and a better results
with low error bars for inactive stars (see the black symbols in
Fig.~\ref{lineratio}).

\begin{figure*}[htb!]
\centering
\includegraphics[angle=0,width=14cm,clip]{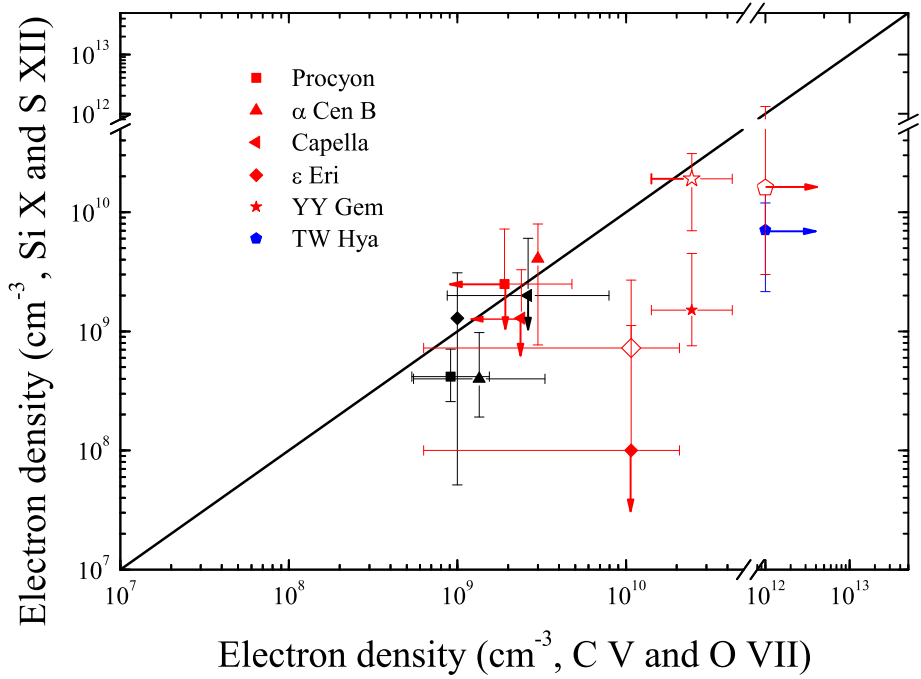}
\caption[short title]{(Color online) Comparisons of the electron
densities derived from K-shell ions (C~V and O~VII) and boron-like
ions (Si~X and S~XII). Black symbols denote results from C~V and
Si~X; red symbols refer to results from O~VII and S~XII. The
linear line indicates the same densities from the K-shell and
boron-like spectra. Symbols without error bars and arrows indicate
no electron density being available. For $\epsilon$~Eri, YY~Gem
and TW~Hya, taking the third-order diffraction into account, the
resulted electron density is shown by red-opened symbols.}
\label{necompar}
\end{figure*}

Using the line fluxes derived in the above section, we diagnose
the electron density for our sample (see the blue symbols in
Fig.~\ref{lineratio}) as listed in Table~\ref{density}. For
Capella, we adopt the line flux from ACIS-S observation, whereas
the line fluxes from the HRC-S observations are used for other
stars in our sample, that is the blending effect has not been
considered. The deduced densities are typical values for inactive
stars. We further compare the densities with the values derived
from spectra of helium-like ions with the same peak line formation
temperatures in the ionization equilibrium~\citep{BBG06} as shown
in Fig.~\ref{necompar}.

The density from Si~X shows a good agreement with the value
constrained by C~V within the statistical errors. For
$\epsilon$~Eri and $\alpha$~Cen~B, electron densities are not
available from the helium-like C~V and O~VII spectra,
respectively. The diagnostic from boron-like Si~X and S~XII
compensates the estimation for the X-ray emitting regions with low
temperatures. For inactive stars and Capella, the density
constrained by S~XII is also in agreement with the results from
O~VII. However, for $\epsilon$~Eri, YY~Gem and TW~Hya, it is clear
that the density constrained by S~XII is lower than the result
from O~VII, which is due to the blending effect from third order
diffraction of the emission line at 12.136~\AA\, resulting from
Fe~XVII and Ne~X. For pre-main sequence star---TW~Hya, \cite{SS04}
estimated the electron density is not less than
1.0$\times10^{12}$cm$^{-3}$, and conclude the X-ray emission being
from the accretion shock.

Adopting the efficiency of third order diffraction determined in
above subsection, and the line flux at 12.136~\AA\,, we extract
its contribution around 36.394~\AA\,, and re-derived the line
ratio and the corresponding electron density as illustrated by
open-symbols in Fig.~\ref{density}. For $\epsilon$~Eri, the
resulted electron density is still lower than the value
constrained by O~VII by an order of magnitude. However, the
determined $n_{\rm e}$ from S~XII shows an agreement with the
$n_{\rm e}$ constrained by O~VII for YY~Gem and TW~Hya within the
1$\sigma$ uncertainty. We also notice that the electron density
determined from S~XII (or O~VII) is slightly higher than that from
Si~X (or C~V) with lower peak line formation temperature.

By using the line ratio of S~XII and Ar~XIV, effective electron
densities of the laboratory plasmas in the Lawrence Livermore EBIT
performed by Lepson and collaborators~\citep{LBB03,LBB05}, are
constrained by the boron-like line ratio as illustrated by the
red-square symbols in Fig.~\ref{lineratio}. For argon plasma, the
constrained electron density
($5.6^{+1.0}_{-1.1}\times10^{10}$~cm$^{-3}$) agrees well with the
actual density (6.0$\times10^{10}$~cm$^{-3}$) estimated by the
K-shell N~VI spectrum as reported by~\cite{CBH04}. For sulphur
plasma, the electron density is firstly estimated to be
$3.4^{+0.8}_{-0.6}\times10^{10}$~cm$^{-3}$. Through the laboratory
measurement, \cite{CBH04} further benchmark the line ratio of
Fe~XXII at the low-density limit. The experimental values are
overlapped again by the red-circle symbols in
Fig.~\ref{lineratio}. The astrophysical ratio (dark-yellow
triangle) of Fe~XXII is derived for the ACIS-S observation of
Capella, and shows the constrained electron density is less than
7.6$\times10^{12}$~cm$^{-3}$, which is consistent with previous
works for the stellar corona.

\section{Conclusions}
Collisional-radiative model calculation reveals that the line
ratio [{\it I}($3d_{5/2}-2p_{3/2}$)+{\it
I}($3d_{3/2}-2p_{3/2}$)]/{\it I}($3d_{3/2}-2p_{1/2}$) of Si~X,
S~XII, Ar~XIV and Fe~XXII is sensitive to the electron density in
the ranges $n_{\rm e}=4.0\times10^7-3.0\times10^{10}$~cm$^{-3}$,
$4.0\times10^8-3.0\times10^{11}$~cm$^{-3}$,
$3.0\times10^9-4.0\times10^{12}$~cm$^{-3}$ and
$2.0\times10^{12}-3.0\times10^{15}$~cm$^{-3}$, respectively. This
compensates the $n_{\rm e}$-diagnostics of K-shell spectra, e.g.
helium-like ions. We further explore the $3d-2p$ transition lines
of boron-like ions in the stellar coronal spectra measured with
the LETG Spectrometer combined with HRC on board the {\it Chandra
X-ray Observatory}. Though the emission lines are very weak, the
electron density constrained from these lines, shows a good
agreement with the those constrained by C~V and O~VII K-shell
spectra inactive stars, and the uncertainties can be comparable
with and even better than those estimated from K-shell spectra.
When the blending effect from third-order diffraction has been
taken into account, the deduced electron density increases, and
gets a better consistency with that estimated from O~VII for
active stars, e.g. YY~Gem. We also notice that the determined
electron density from S~XII is higher than that constrained from
Si~X, which is due to its higher peak temperature of line
formations in collisional equilibrium.

By using the line ratio, effective electron densities in the
electron beam ion trap (EBIT) plasma performed
by~\cite{LBB03,LBB05} at Lawrence Livermore EBIT, are constrained
to be $n_{\rm e}=3.4^{+0.8}_{-0.6}\times10^{10}$~cm$^{-3}$ and
$5.6^{+1.0}_{-1.1}\times10^{10}$~cm$^{-3}$ for sulphur and argon
plasmas. In case of argon, a good agreement is found with the
actual electron density derived from N~VI K-shell spectrum.

In conclusion, the boron-like $3d-2p$ spectra provides a good
$n_{\rm e}$-diagnostics for hot plasmas, and compensates the
spectral diagnostic of K-shell spectrum.

\begin{acknowledgements}
G.Y. thanks Prof. Badnell, University of Strathclyde in United
Kingdom, for his constructive suggestions. This study is supported
by the National Natural Science Foundation of China under Grant
Nos. 10603007, 10521001 and 10573024, as well as the National
Basic Research Program of China (973 program) under grant
No.~2007CB815103.
\end{acknowledgements}

\begin{table*}
\centering
    \caption[I]{Wavelengths of three interested $3d$--$2p$ transitions of Si~X, S~XII, Ar~XIV and Fe~XXII.}\label{wavelength}
    \vspace{0.2cm}
      \[
      \begin{array}{c|cccc} \hline\hline
 {\rm Transitions}   & \multicolumn{4}{c}{\rm Wavelength~(\AA\,)}
 \\ \cline{2-5}
                 & {\rm Si~X} & {\rm S~XII} & {\rm Ar~XIV} & {\rm Fe~XXII} \\
 \hline
2s^23d~^2D_{3/2}-2s^22p~^2P_{1/2} & 50.524 & 36.398 & 27.469 & 11.769 \\
2s^23d~^2D_{5/2}-2s^22p~^2P_{3/2} & 50.691 & 36.564 & 27.629 & 11.921 \\
2s^23d~^2D_{3/2}-2s^22p~^2P_{3/2} & 50.703 & 36.573 & 27.642 & 11.936 \\
\hline
         \end{array}
      \]
      \end{table*}

\begin{table*}
\centering
    \caption[I]{Summary of stellar properties and X-ray
    luminosity (5---175~\AA\,) for the stars.}\label{stars}
    \vspace{0.2cm}
      \[ \hspace{-2.0cm}
      \begin{array}{lcccccccc} \hline\hline
{\rm star} & {\rm ObsID} & {\rm t^a_{obs}} & {\rm Spectral Type} &
{\rm distance^b} & T{\rm ^a_{eff}} & {\rm log({\it L}_{bol})} &
{\rm R_{\star}^b} & L{\rm _X^c} \\
  &    & {\rm ks} &  & {\rm pc} & {\rm K} & {\rm erg/s} & {\rm [{\it R}_{\sun}]} &
  10^{28}{\rm erg/s} \\ \hline
{\rm Procyon}      & 63+1224+1461 & 70.15+20.93+70.25 & {\rm F5.01V-V}      & 3.5   & 6540  & 34.46 & 2.06     & 2.43 \\
{\rm \alpha~Cen~B} & 29           & 79.5              & {\rm K0.0V}         & 1.34  & 5780  & 33.28 & 0.8      & 0.52 \\
{\rm Capella}      & 1248/55      & 84.7/53.48        & {\rm G1.0III/K0.0I} & 12.94 & 5850  & 35.71 & 9.2/13   & 255  \\
{\rm \epsilon~Eri} & 1869         & 105.3             & {\rm K2.0V}         & 3.22  & 4780  & 33.11 & 0.81     & 20.9 \\
{\rm YY~Gem}       & 28           & 57.63             & {\rm dMIe/dMIe}     & 14.7  &       &       &0.66/0.58 & 54.4 \\
{\rm TW~Hya}       & 6443         & 150.24            & {\rm K8V}           & 56.0  &       &       &          & 150.0^d\\
\hline
         \end{array}
      \]
\flushleft{$^a$ Exposure time of observation \newline $^b$
From~\cite{NGS04}
\newline $^c$ From~\cite{NSB02a} \newline $^d$ Derived from {\it XMM-}Newton observation~\citep{SS04}.}
      \end{table*}

\begin{table*}
\centering
    \caption[I]{Measured flux (in unit of $1.0\times10^{-4}$~photon~s$^{-1}$~cm$^{-2}$) of S XII lines
    at 36.398~\AA\, and 36.564~\AA\, and the line at 12.136~\AA\, for 6 stars observed with the LETG spectrometer. $F^-$ and $F^+$
    indicate the line fluxes with 1$\sigma$ error derived from negative and positive diffraction spectra, respectively.}
    \vspace{0.2cm}\label{lineflux}
      \[
      \begin{array}{lcccccc} \hline\hline
{\rm Stars} & {\rm Instrument} & F^-{\rm (36.398\AA)} & F^+{\rm (36.398\AA)} & F^-{\rm (36.564\AA)} & F^+{\rm (36.564\AA)} & F{\rm (12.136\AA)} \\
\hline
{\rm Procyon     } & {\rm HRC-S} & 0.38\pm0.11 & 0.33\pm0.13 & 0.24\pm0.11  & 0.32\pm0.13 &      -      \\
{\rm Procyon^a   } & {\rm HRC-S} & 0.34\pm0.14 &     -       & 0.24\pm0.13  &     -       &      -      \\
{\rm Procyon^a   } & {\rm RGS1 } & 0.35\pm0.10 &     -       & 0.15\pm0.06  &     -       &      -      \\
{\rm Procyon^a   } & {\rm RGS2 } & 0.28\pm0.07 &     -       & 0.29\pm0.09  &     -       &      -      \\
{\rm \alpha~Cen~B} & {\rm HRC-S} & 0.43\pm0.11 &     -       & 0.32\pm0.11  &     -       &      -      \\
{\rm Capella     } & {\rm HRC-S} & 2.61\pm0.89 & 2.84\pm0.87 & 0.89\pm0.23  & 0.91\pm0.24 &11.04\pm0.30 \\
{\rm Capella     } & {\rm ACIS } & 0.69\pm0.14 &     -       & 0.34\pm0.14  &     -       &      -      \\
{\rm \epsilon~Eri} & {\rm HRC-S} & 0.71\pm0.13 & 0.78\pm0.15 & 0.23\pm0.11  & 0.45\pm0.13 & 0.91\pm0.04 \\
{\rm YY~Gem      } & {\rm HRC-S} & 0.29\pm0.06 & 0.49\pm0.09 & 0.15\pm0.07  & 0.19\pm0.09 & 1.03\pm0.07 \\
{\rm TW~Hya      } & {\rm HRC-S} & 0.37\pm0.10 & 0.42\pm0.11 & 0.36\pm0.10  & 0.13\pm0.09 & 0.48\pm0.03 \\
\hline
         \end{array}
      \]
      \flushleft{$^{\rm a}$ The line fluxes in this case are from Raassen et al. (2002).}
      \end{table*}
\clearpage

\begin{table*}
\centering
    \caption[I]{Diagnosed electron densities (in unit of cm$^{-3}$) from Si X and S XII for our sample.}
    \vspace{0.2cm}\label{density}
      \[
      \begin{array}{lcc} \hline\hline
{\rm Stars} & n_{\rm e}{\rm (Si~X)} &  n_{\rm e}{\rm (S~XII)} \\
\hline
{\rm Procyon     } & 4.2_{-1.6}^{+2.9}\times10^8 &  2.5^{+4.7}\times10^9         \\
{\rm \alpha~Cen~B} & 4.0_{-2.1}^{+5.8}\times10^8 &  4.1_{-3.3}^{+3.9}\times10^9  \\
{\rm Capella     } & 2.0^{+4.0}\times10^9        &  1.3^{+2.0}\times10^9         \\
{\rm \epsilon~Eri} & 1.3_{-1.2}^{1.8}\times10^9  &  1.0^{+10.2}\times10^8        \\
{\rm YY~Gem      } & -                           &  1.5_{-0.8}^{+3.0}\times10^9  \\
{\rm TW~Hya      } & -                           &  7.1_{-4.9}^{+9.1}\times10^9  \\
\hline
         \end{array}
      \]
          \end{table*}

\clearpage

\newpage
\clearpage
\newpage
\clearpage

\newpage
\clearpage

\newpage
\clearpage

\end{document}